\documentstyle[12pt,epsf]{article}
\begin{document}

\begin{flushright}
WUE-ITP-2000-015\\[-0.15cm]
hep-ph/0006327\\[-0.15cm]
June 2000
\end{flushright}

\begin{center}
{\Large {\bf Search for solar Kaluza-Klein axions}}\\[0.3cm]
{\Large {\bf in theories of low-scale quantum gravity }}\\
\vskip1.2cm

{\large L. Di Lella$^a$, A. Pilaftsis$^b$, G. Raffelt$^{c,d}$ and 
             K. Zioutas$^{a,e}$}\\[0.35cm]
$^a${\em CERN, CH-1211 Geneva 23, Switzerland}\\[0.2cm]
$^b${\em Institut f\"ur Theoretische Physik, Universit\"at W\"urzburg,\\
Am Hubland, 97074 W\"urzburg, Germany}\\[0.2cm]
$^c${\em Max-Planck-Institut f\"ur Physik, F\"ohringer Ring 6, 80805
 Munich, Germany}\\[0.2cm]
$^d${\em TECHNION (Israel Institute of Technology), Haifa 32000, 
Israel}\\[0.2cm]
$^e${\em Department of Physics, University of Thessaloniki,}\\
{\em GR 54006 Thessaloniki, Greece}
\end{center}
\vskip1.cm \centerline{\bf ABSTRACT}  
We  explore  the  physics  potential  of a  terrestrial  detector  for
observing axionic Kaluza-Klein excitations  coming from the Sun within
the  context  of  higher-dimensional  theories  of  low-scale  quantum
gravity.   In  these theories,  the  heavier  Kaluza-Klein axions  are
relatively  short-lived   and  may  be  detected   by  a  coincidental
triggering of  their two-photon decay  mode.  Because of  the expected
high  multiplicity   of  the   solar  axionic  excitations,   we  find
experimental sensitivity  to a fundamental Peccei-Quinn  axion mass up
to $10^{-2}$~eV  (corresponding to an  effective axion-photon coupling
$g_{a\gamma \gamma} \approx 2.\times 10^{-12}$~GeV$^{-1}$) in theories
with  2  extra  dimensions  and a  fundamental  quantum-gravity  scale
$M_{\rm  F}$  of  order  100~TeV,  and  up  to  $3.\times  10^{-3}$~eV
(corresponding    to     $g_{a\gamma    \gamma}    \approx    6.\times
10^{-13}$~GeV$^{-1}$) in theories with  3 extra dimensions and $M_{\rm
  F}=1$  TeV.  For  comparison,  based on  recent  data obtained  from
lowest  level  underground  experiments,  we derive  the  experimental
limits:   $g_{a   \gamma   \gamma}   \stackrel{<}{{}_\sim}   2.5\times
10^{-11}$~GeV$^{-1}$  and $g_{a  \gamma  \gamma} \stackrel{<}{{}_\sim}
1.2\times 10^{-11}$~GeV$^{-1}$  in the aforementioned  theories with 2
and 3 large compact dimensions, respectively.
  
\newpage

\setcounter{equation}{0}
\section{Introduction}

In  superstring theories it  turns  out  to be  possible to  lower the
string  scale       without     lowering    the                 Planck
scale~\cite{IA,Witten,JL,ADD,DDG0}.     Most   notably,  Arkani-Hamed,
Dimopoulos and Dvali \cite{ADD} have  proposed the radical possibility
that  the fundamental  scale of  quantum  gravity  might no longer  be
associated  with the Planck mass $M_{\rm  P} = 1.2\times 10^{19}$ GeV,
but  the true  scale  of quantum gravity,  $M_{\rm  F}$, could be many
orders of  magnitude smaller than $M_{\rm  P}$, close to TeV energies. 
In  such a    novel theoretical  framework, the  standard-model   (SM)
particles can only live  in  a $(1+3)$-dimensional Minkowski  subspace
that constitutes   our observable world,   whereas gravity  may freely
propagate to a number $n$ of large extra dimensions.  Furthermore, the
ordinary Planck  mass $M_{\rm P}$  would be  related to the  genuinely
fundamental scale $M_{\rm F}$ through
\begin{equation}
  \label{Gauss} M_{\rm P}\ \approx\ M_F\, (R\, M_{\rm F})^{n/2}\, ,
\end{equation} 
where $R$ denotes the compactification  radius, which is considered to
be common   for all extra compact  dimensions.   The case $n =  1$ and
$M_{\rm   F}$  of  order    TeV   leads  to  a   visible   macroscopic
compactification  radius and  is   therefore  not  viable.   Moreover,
astrophysical and cosmological  considerations  give rise to  a  lower
limit on $M_{\rm F}$ of  order 100 TeV, for  the scenario with $n = 2$
extra dimensions \cite{HS}, while  $M_{\rm F}$ can be  as low as 1 TeV
for theories with $n > 2$ dimensions.

In addition to gravity, one might think that fields which are singlets
under    the Standard-Model gauge group  could   also propagate in the
$[1+(3+n)]$-dimensional space.  As such, one might consider isosinglet
neutrinos \cite{ADDM,DDG,AP}  or  axion fields \cite{ADD,CTY,DDG1}. In
fact, within the context of theories of TeV-scale quantum gravity, the
latter realization is theoretically compelling for the solution of the
strong  CP problem through the  Peccei-Quinn (PQ) mechanism. According
to this idea, the strong  CP-odd parameter $\theta$ may be dynamically
eliminated  by means  of the  spontaneous  breakdown of  a global U(1)
symmetry.   On the    other hand,  phenomenological  and astrophysical
considerations  place  lower and upper  limits  on the  breaking scale
$v_{\rm PQ}$ of the PQ-U(1) symmetry,  which has to  be many orders of
magnitude larger than the TeV scale of quantum gravity.  Therefore, in
order  to account for   this large mass  scale,  one inevitably has to
introduce   a  singlet  higher-dimensional axion field    into the QCD
Lagrangian,  with a higher-dimensional PQ-breaking scale $\bar{v}_{\rm
  PQ}$  that could even be  much smaller than 1  TeV.  As  we will see
below,  as  a   result of   the  compactification of  the large  extra
dimensions, the  effective four-dimensional  PQ-breaking scale $v_{\rm
  PQ}$ can be obtained from  $\bar{v}_{\rm PQ}$, after multiplying the
latter  by the  huge   higher-dimensional  volume factor  $(M_{\rm  F}
R)^{n/2} \approx M_{\rm P}/M_{\rm  F}$.  In this way, the  PQ-breaking
scale   $v_{\rm  PQ}$  may  reside in  the  phenomenologically allowed
region.  Another feature of the higher-dimensional axionic theories is
that their mass  spectrum  consists of a  tower  of  Kaluza-Klein (KK)
excitations,  which have an almost  equidistant  mass-spacing of order
$1/R$.   The lowest KK excitation may  be identified with the ordinary
PQ axion and specifies the strength of each KK state to matter.

This tower  of axionic  modes has  two phenomenological  consequences. 
First, for a fixed value  of the axion coupling  constant to matter or
photons, a given source such as the Sun will emit  axions of each mode
up to the kinematic limit. The high multiplicity of the KK axion modes
thus leads  to  much larger flux   than would be otherwise  expected.  
Second,   the large mass  of the  KK modes   compared  to usual axions
dramatically   increases      the  width   of     the  decay   process
$a\to\gamma\gamma$  by  opening up phase   space.  Therefore,  one may
plausibly search for the decay photons of the solar KK axion flux in a
laboratory experiment.

In this paper,  we shall analyze  the potential of a terrestrial axion
detector to observe the  radiative decay of solar KK  axion modes.  In
particular, such a detector proves inexpensive and may run in parallel
with the CERN Axion Solar Telescope (CAST) which  will be built from a
decommissioned LHC  test  magnet \cite{CAST}.  We  will  find that the
suggested terrestrial detector may reach the unprecedented sensitivity
of the $10^{-2}$-eV    level ($g_{a\gamma \gamma}   \approx  10^{-12}$
GeV$^{-1}$) to the fundamental PQ axion mass ($m_{\rm PQ}$).

The paper  is organized as  follows: in Section  2 we briefly describe
the basic  low-energy   structure of a  generic   theory that includes
higher-dimensional axions.  In Section 3 we compute  the solar flux of
massive KK axions. In Section 4 we estimate the event rates of photons
due to axion  decays  as seen  by  a terrestrial detector.  Section  5
summarizes our conclusions.

\setcounter{equation}{0}
\section{Axions in large extra dimensions}

Before discussing the higher-dimensional case, let us first recall the
main  phenomenological  predictions  of the   axion  theories in  four
dimensions. The axionic sector of the effective Lagrangian which is of
interest to us has the generic form
\begin{equation}
  \label{Laxion}
{\cal L}_{\rm eff}\ =\ 
\frac{1}{2}\, (\partial_\mu a)(\partial^\mu a)\: -\: \frac{1}{2}\,
m^2_{\rm PQ}\,a^2\: +\:  \frac{g_{a\gamma\gamma}}{4}\, a\, 
F_{\mu\nu}\, \widetilde{F}^{\mu\nu}\, ,
\end{equation}
where  $a$ is the PQ  axion, $F_{\mu\nu}$ and $\widetilde{F}^{\mu\nu}$
are the electromagnetic field-strength tensor  and its associate  dual
tensor, and
\begin{equation}
  \label{axgam}
g_{a\gamma\gamma}\ =\ \frac{\xi\,\alpha_{\rm em}}{\pi}\
\frac{1}{v_{\rm PQ}} 
\end{equation}
is the   effective  axion-photon-photon coupling.   The multiplicative
parameter $\xi$ in Eq.\ (\ref{axgam}) is generally of order unity, and
crucially depends on  the  axion model under  study  \cite{ZDFS,KSVZ}.
Furthermore, the  PQ  axion mass $m_{\rm  PQ}$  is  related to the
breaking scale $v_{\rm PQ}$ of the PQ U(1) symmetry through
\begin{equation}
  \label{vPQ}
m_{\rm PQ}\ \sim\ \frac{m^2_\pi}{v_{\rm PQ}}\ ,
\end{equation}
where $m_\pi \approx 135$ MeV is the pion mass.  
Astrophysical and cosmological limits~\cite{GR99} 
indicate that
\begin{equation}
  \label{vPQconstr}
10^9\ {\rm GeV}\ \stackrel{<}{{}_\sim}\ v_{\rm PQ}\ \stackrel{<}{{}_\sim}\
10^{12}\ {\rm GeV}\, ,
\end{equation}
which in turn by virtue of Eq.\ (\ref{vPQ}) implies that
\begin{equation}
  \label{mPQconstr}
10^{-2}\ {\rm eV}\ \stackrel{>}{{}_\sim}\ m_{\rm PQ}\ \stackrel{>}{{}_\sim}\
10^{-5}\ {\rm eV,}\
\end{equation}
respectively. The lifetime of the PQ axion is easily calculated to be
\begin{equation}
  \label{taugg}
\tau ( a\to \gamma \gamma)\ =\ 
\frac{64\pi }{g^2_{a\gamma\gamma} m^3_{\rm PQ}}\
\approx\ 10^{48}\ {\rm days}
\,\times \bigg(\frac{10^{-15}\ {\rm GeV}^{-1}}{g_{a\gamma\gamma}}\bigg)^2\
\bigg(\frac{10^{-5}\ {\rm eV}}{m_{\rm PQ}}\bigg)^3     \, .
\end{equation}
For $g_{a\gamma\gamma} =   10^{-15}$ GeV$^{-1}$, which corresponds  to
$m_{\rm PQ}  = 10^{-5}$ eV,  the axion lifetime turns   out to be much
larger than the   age of the    universe.  The prospect  of  detecting
photonic  axion decays would  have remained hopeless,  even if one had
considered larger  axion masses.    For instance,  for $m_{\rm PQ}   =
10^{-1}$ eV ($g_{a\gamma\gamma} =   10^{-11}$ GeV$^{-1}$),  the  axion
decay is still undetectable with  a lifetime $\tau (a\to \gamma\gamma)
\approx 10^{27}$ days.

We  shall  now  focus   on  the  higher-dimensional  case.   Following
Refs.~\cite{ADD,CTY,DDG1},  we  introduce   one  singlet  axion  field
$a(x^\mu, {\bf y})$  which feels the presence of  a number $\delta \le
n$   of   large   extra    dimensions,   denoted   by   ${\bf   y}   =
(y_1,y_2,\dots,y_\delta)$.   The relevant axionic  sector may  then be
determined by the effective Lagrangian
\begin{equation}
  \label{Lhigher} 
{\cal L}_{\rm eff}\ =\ \int d^\delta {\bf y}\, \bigg[\, \frac{1}{2}\, 
M^\delta_F\, (\partial_\mu a)(\partial^\mu a)\: +\:
\frac{1}{2}\, M^\delta_F\, (\partial_\delta a)(\partial^\delta a)\: +\:
\delta^{(\delta )} ({\bf y})\, 
\frac{\xi\,\alpha_{\rm em}}{\pi}\, \frac{a}{\bar{v}_{\rm PQ}}\,
F_{\mu\nu}\, \widetilde{F}^{\mu\nu}\, \bigg]\, ,
\end{equation}
where  $\bar{v}_{\rm  PQ}$  denotes  the  original  higher-dimensional
PQ-breaking  scale.   In  Eq.\  (\ref{Lhigher}), the  axion  field  is
compactified  on  a  {\bf  Z}$_2$  orbifold with  an  orbifold  action
\cite{DDG1}: ${\bf y}  \to -{\bf y}$, i.e.\ the  axion field satisfies
the properties: $a(x^\mu, {\bf y} ) = a (x^\mu, {\bf y} + 2\pi R)$ and
$a(x^\mu, {\bf y})  = a (x^\mu, -{\bf y})$.  The  latter gives rise to
the KK decomposition:
\begin{equation}
  \label{KKdec}
a( x^\mu, {\bf y})\ =\ \sum_{\bf n=0}^\infty\, a_{\bf n} (x^\mu)\, \cos\bigg( 
\frac{{\bf n\, y}}{R}\bigg)\ ,
\end{equation}
where ${\bf  n} = (n_1,n_2,\dots,n_\delta)$  is a $\delta$-dimensional
vector  that  labels the  individual  KK  excitations, and  $\sum_{\bf
  n=0}^\infty  \equiv  \sum_{n_1=0}^\infty  \sum_{n_2=0}^\infty  \dots
\sum_{n_\delta=0}^\infty$.  Substituting  Eq.\ (\ref{KKdec}) into Eq.\ 
(\ref{Lhigher})  and taking  the PQ  mechanism into  consideration, we
arrive at the effective Lagrangian \cite{DDG1}
\begin{equation}
  \label{Ldelta} 
{\cal L}_{\rm eff}\ =\ \frac{1}{2}\, \sum_{\bf n=0}^\infty\, (\partial_\mu
a_{\bf n})(\partial^\mu a_{\bf n})\: -\: \frac{1}{2}\, m^2_{\rm PQ}\,a_0^2\:
-\: \frac{1}{2}\, \sum_{\bf n\neq 0}^\infty\, \frac{{\bf n}^2}{R^2}\ 
a_{\bf n}^2\: +\:
\frac{\xi\,\alpha_{\rm em}}{\pi}\, \sum_{\bf n=0}^\infty\, 
\frac{r_{\bf n} a_{\bf n}}{v_{\rm PQ}}\, 
F_{\mu\nu}\, \widetilde{F}^{\mu\nu}\, ,
\end{equation}
with  $r_0  =  1$  and  $r_{\bf  n\neq 0} =  \sqrt{2}$.  {}From  Eq.\ 
(\ref{Ldelta}), it is easy to  read off the effective couplings of the
KK axions to photons,
\begin{equation}
  \label{axgam1}
g_{a_{\bf n}\gamma\gamma}\ =\ \frac{r_{\bf n}\xi\,\alpha_{\rm em}}{\pi}\
\frac{1}{v_{\rm PQ}}\ \approx\ g_{a\gamma\gamma}\, .
\end{equation}
Instead  of a {\bf  Z}$_2$ orbifold  compactification, one  could have
equally  considered  the  compactification on  a  $\delta$-dimensional
torus~\cite{ADD}, leading to modified coupling constants by factors of
order unity.   For the sake of  simplicity we will  always assume that
the  KK axion  modes  couple to  photons  with the  usual PQ  coupling
$g_{a\gamma\gamma}$; it  is trivial to  insert model-dependent factors
in the final result.

Few  comments are now  in order  in connection  with the  effective KK
Lagrangian   (\ref{Ldelta}).   First,  we   should  remark   that  the
higher-dimensional  PQ-breaking scale $\bar{v}_{\rm  PQ}$ may  be very
low at the TeV scale,  when compared to the usual four-dimensional one
$v_{\rm PQ}$, i.e.\ 
\begin{equation}
  \label{vPQrel}
\bar{v}_{\rm PQ}\ \approx\ \bigg(
\frac{M_{\rm F}}{M_{\rm P}}\bigg)^{\delta/n}\ v_{\rm PQ}\, . 
\end{equation}
The suppression mechanism is very  analogous to the case, in which the
fundamental scale of quantum gravity can be reduced to the electroweak
scale in the presence of  large extra dimensions \cite{ADD} (cf.\ Eq.\ 
(\ref{Gauss})).  In  Eq.\ (\ref{vPQrel}),  the simplest setting  is to
consider  that   both  gravity  and   axions  live  within   the  same
higher-dimensional  space, i.e.\  $\delta =  n$.  Second,  one notices
that the lowest KK state constitutes  the PQ axion of the theory which
determines  the size  of the  coupling of  the KK  axions to  photons. 
Finally, the KK-axion masses are given by
\begin{equation}
  \label{maxion}
m_{a_0}\ =\ m_{\rm PQ}\ \ll \frac{1}{R}\, ,\qquad
m_{a_{\bf n}}\ \approx\ \frac{n}{R}\, ,
\end{equation}
with $n  = |{\bf n}| = \sqrt{n^2_1  + \cdots + n^2_\delta}>  0$. It is
interesting to observe that for the higher-dimensional scenarios under
discussion, the mass-spacing of the KK axions is always larger than PQ
masses lying  in the phenomenologically favoured  region, with $m_{\rm
  PQ}\stackrel{<}{{}_\sim} 0.01$ eV. For example, for $\delta = 2$ and
$M_{\rm F}  \approx 100$ TeV \cite{HS},  one obtains $1/R  \sim 1$ eV,
while for $\delta  = 3$ and $M_{\rm F} \approx 1$  TeV, the inverse of
the compactification  radius reaches a  much higher value,  i.e.\ $1/R
\sim 10$~eV.

The lifetime of an individual  axionic KK state $a_{\bf n}$ may easily
be computed from Eq.~(\ref{taugg}). In this way, we find
\begin{equation}
  \label{tauxn}      
\tau  (    a_{\bf n}\to  \gamma   \gamma    )\ \approx\
\bigg(\frac{m_{\rm PQ}}{m_{a_{\bf n}} }\bigg)^3\,    
\tau  (a_0\to   \gamma\gamma)\,  .    
\end{equation}  
We observe  that the  lifetime of the  KK axion $a_{\bf  n}$ decreases
rapidly with the third power of its mass. For example, the lifetime of
one single  (solar) KK-axion  mode with $m_{a_{\bf  n}} =  10$~keV and
$g_{a\gamma\gamma}  = 10^{-11}$~GeV$^{-1}$  (corresponding  to $m_{\rm
  PQ} = 10^{-1}$~eV)  is $\tau ( a_{\bf n}\to  \gamma \gamma ) \approx
10^{12}$~days,  which  is 15  orders  of  magnitude  smaller than  the
respective  one  obtained in  usual  four-dimensional  theories of  PQ
axions.

\setcounter{equation}{0}
\section{Solar flux of Kaluza-Klein axions}

\subsection{Primakoff process}

In order to calculate the solar flux of KK axion modes we restrict
ourselves to hadronic axion models where these particles do not couple
to electrons at tree level.  The dominant production processes will
thus involve the axion-photon interaction; the axion-nucleon coupling
will not be important in the Sun.  The usual PQ axions are primarily
produced by the Primakoff process $\gamma+Ze\to Ze+a$ where a thermal
photon in the solar interior converts into an axion in the Coulomb
fields of nuclei and electrons in the solar plasma.  In addition, the
KK modes can be produced by the photon coalescence process
$\gamma\gamma\to a$. For PQ axions, this process is suppressed by the
small mass and actually is kinematically forbidden in the solar plasma
because the effective photon mass (plasma frequency) is about 0.3~keV.
However, with a temperature in the Sun of around 1.3~keV, the solar KK
axions will be produced with masses up to several keV, rendering the
coalescence process an important contribution.

Beginning with the Primakoff process, the production cross section on
a target with charge $Ze$ in a nonrelativistic plasma is found to
be~\cite{Raffelt86}
\begin{equation}
\frac{d\sigma_{\gamma\to a}}{d\Omega}=
\frac{g_{a\gamma\gamma}^2Z^2 \alpha}{8\pi}\,
\frac{|{\bf k}\times{\bf p}|^2}{{\bf q}^4}\,
\frac{{\bf q}^2}{{\bf q}^2+\kappa^2}\ ,
\end{equation}
where ${\bf k}$ is the photon  momentum, ${\bf p}$ the axion momentum,
and ${\bf q}={\bf k}-{\bf p}$ the  momentum transfer.  The last factor
takes account of  screening effects where the Debye-H\"uckel screening
scale is given by
\begin{equation}
  \label{kappa}
\kappa^2=\frac{4\pi\alpha}{T}\,\frac{\rho}{m_u}
\left(Y_e+\sum_j Z_j^2 Y_j\right)\, .
\end{equation}
In Eq.\ (\ref{kappa}),  $\rho$ is the mass  density,  $m_u$ the atomic
mass   unit  (approximately the  proton   mass),  $Y_e$ the  number of
electrons per baryon  in the medium, and  $Y_j$ the number of  various
nuclear species $j$ per baryon with nuclear  charge $Z_j$.  The medium
is assumed to be nonrelativistic,  and  recoil effects by the  targets
have been neglected since typical  photon  energies of  a few keV  are
much smaller  than even the electron mass.   It turns out that we have
the approximate relation,  $\kappa\approx 7 T$, between the  screening
scale $\kappa$ and the temperature $T$ in  the relevant regions of the
Sun.

Summing over all   target   species of the medium,    the photon-axion
transition rate is finally
\begin{equation}
\Gamma_{\gamma\to a}\ =\ \frac{g_{a\gamma\gamma}^2 T \kappa^2}{32\pi^2}\,
\frac{|{\bf k}|}{\omega}\int d\Omega\,
\frac{|{\bf k}\times{\bf p}|^2}{{\bf q}^2({\bf q}^2+\kappa^2)}\, ,
\end{equation}
where $\omega$ is the photon energy  and the factor $|{\bf k}|/\omega$
is   the relative velocity between  photons  and target particles. The
angular integration can be performed explicitly, leading to
\begin{eqnarray}
\Gamma_{\gamma\to a}&=&\frac{g_{a\gamma\gamma}^2 T \kappa^2}{32\pi}\,
\frac{k}{\omega}\,
\Biggl\{\frac{\left[(k+p)^2+\kappa^2\right]\,
\left[(k-p)^2+\kappa^2\right]}{4\,k\,p\,\kappa^2}
\,\ln\left[\frac{(k+p)^2+\kappa^2}{(k-p)^2+\kappa^2}\right]
\nonumber\\
&&\hskip5em{}
-\frac{(k^2-p^2)^2}{4\,k\,p\,\kappa^2}
\ln\left[\frac{(k+p)^2}{(k-p)^2}\right]
-1\Biggr\}\, ,
\end{eqnarray}
where $k=|{\bf k}|$ and $p=|{\bf p}|$.

The effective ``photon mass'' in  the medium, the plasma frequency, is
small in the Sun, typically  about 0.3~keV, while the temperature near
the solar center is $T=1.3~{\rm keV}$  and typical photon energies are
$3  T\approx 4~{\rm keV}$.  Therefore, we ignore  the plasma frequency
and treat photons  as strictly massless.  In a photon-axion transition
the  energy is conserved because  we ignore recoil effects. Therefore,
we use $k=E$ with $E$ the  axion energy and $p=\sqrt{E^2-m^2}$ so that
finally
\begin{equation}
\Gamma_{\gamma\to a}=\frac{g_{a\gamma\gamma}^2 T \kappa^2}{32\pi}\,
\Biggl\{\frac{(m^2-\kappa^2)^2+4E^2\kappa^2}{4\,E\,p\,\kappa^2}
\,\ln\left[\frac{(E+p)^2+\kappa^2}{(E-p)^2+\kappa^2}\right]
-\frac{m^4}{4\,E\,p\,\kappa^2}
\ln\left[\frac{(E+p)^2}{(E-p)^2}\right]
-1\Biggr\}\, .
\end{equation}
Note that  the expression in  curly brackets expands for small momenta
as
\begin{equation}
\{\ldots\}=\frac{8 p^2}{3(\kappa^2+m^2)}+{\cal O}(p^4)\, ,
\end{equation}
so that the emission of slow-moving axions is suppressed.

The axion flux at Earth, differential with regard to the axion energy
$E$, is then found by multiplying the transition rate with the
blackbody photon flux in the Sun, and integrating over a standard
solar model,
\begin{equation}
\Phi_a\ =\ \frac{dF_a}{dE}\ =\ 
\frac{1}{4\pi d_\odot^2}
\int_{\rm sun} d^3{\bf r}\,\,\Gamma_{\gamma\to a}\,
\frac{1}{\pi^2}\,\frac{E^2}{e^{E/T}-1}\ .
\end{equation}
Here  $T$ and   $\kappa^2$ depend on   the  location in the  Sun  and
$d_\odot=1.50\times 10^{13}~{\rm cm}$ is the distance  to the Sun.  We
stress that no velocity factor appears for massive axions because in a
stationary situation all  axions produced per  second  must traverse a
spherical shell around the Sun within one second.

\begin{figure}[p]
   \leavevmode
 \begin{center}
   \epsfxsize=16.0cm
    \epsffile[0 0 539 652]{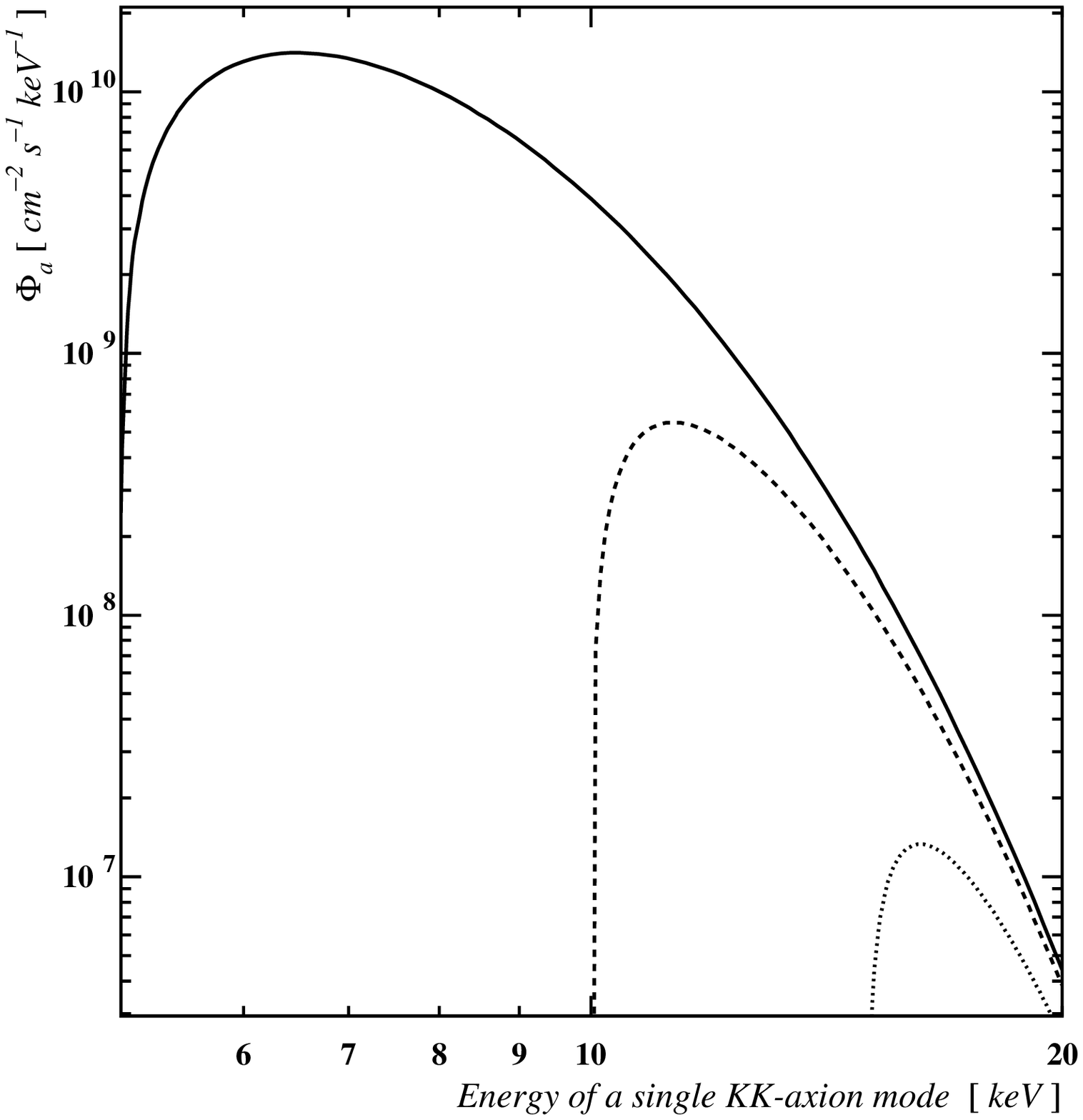}
 \end{center}
 \vspace{-0.5cm} 
\caption{Energy dependence of the solar flux of KK axions at a distance of 
  1~AU due  to Primakoff  process, assuming  a KK axion  mass $m  = 5$
  (solid  line),  10  (dashed  line)  and 15  (dotted  line)~keV,  and
  $g_{a\gamma\gamma} = 10^{-10}$ GeV$^{-1}$.}
\label{f2}
\end{figure}

In Ref.~\cite{Raffelt89} an  approximation formula for  the axion flux
at Earth was given which we slightly modify  and extend to the case of
massive KK axions,
\begin{equation}
\Phi_a\ =\ 
4.20\times10^{10}~{\rm cm^{-2}~s^{-1}~keV^{-1}}
\, \bigg(\frac{g_{a\gamma\gamma}}{10^{-10}\ {\rm GeV}^{-1}}\bigg)^2\,
\frac{E\,p^2}{e^{E/1.1}-0.7}
\,(1+0.02\,m)\, ,
\end{equation}
where $E$, $p$  and $m$ are to be measured  in keV. This approximation
formula is typically  good to better than $\pm  15\%$ for all relevant
conditions, and even  better than a few percent  for the most relevant
case of  axion masses of  larger than a  few keV. In  Fig.~\ref{f2} we
show the energy  dependence of the flux of massive  KK axions at Earth
for three typical choices of the axion mass: $m= 5$, 10 and 15~keV.

\subsection{Photon Coalescence}

\begin{figure}[p]
   \leavevmode
 \begin{center}
   \epsfxsize=16.0cm
    \epsffile[0 0 539 652]{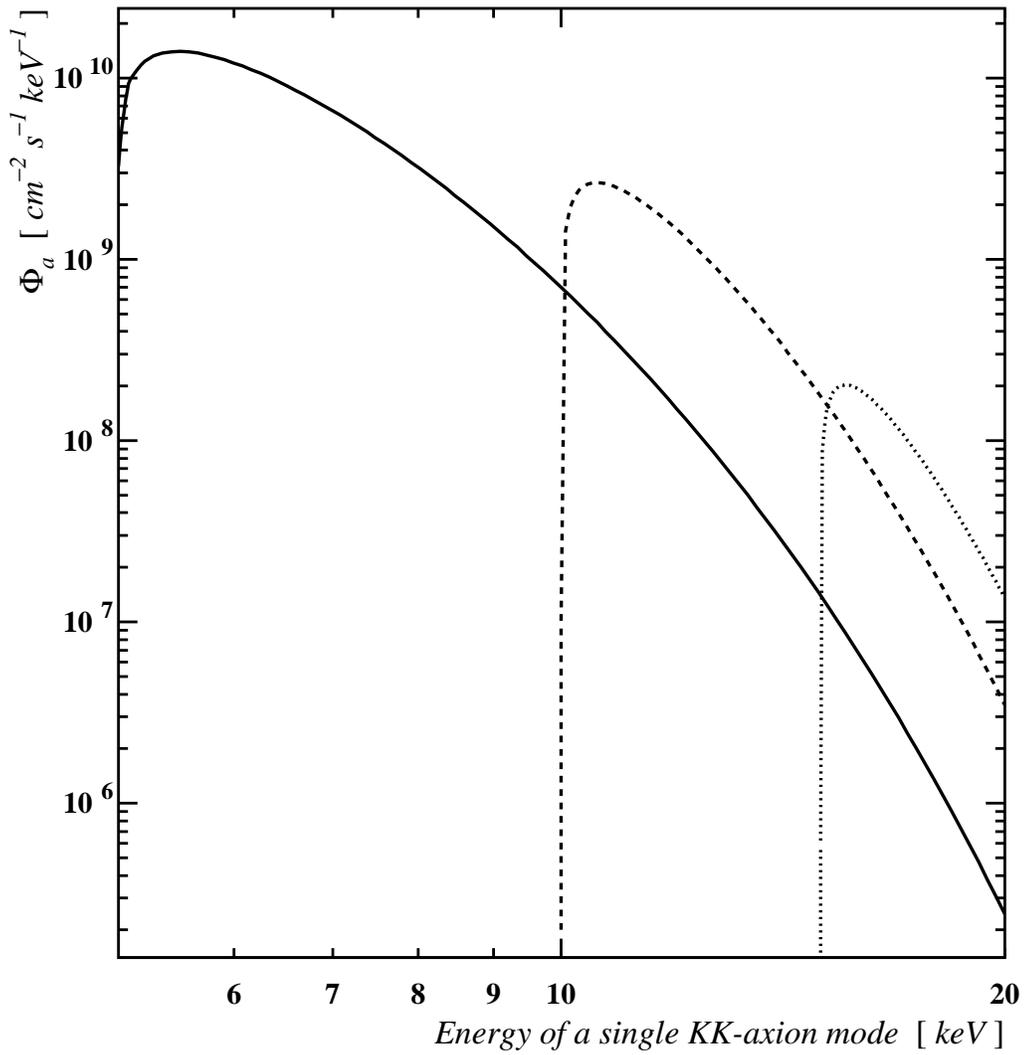}
 \end{center}
 \vspace{-0.7cm} 
\caption{Energy dependence of the solar flux of KK axions at a
  distance of 1~AU due to photon coalescence, assuming a KK axion mass
  $m = 5$ (solid line), 10 (dashed line) and 15 (dotted line)~keV, and
  $g_{a\gamma\gamma} = 10^{-10}$ GeV$^{-1}$.}
\label{f3}
\end{figure}

In order to  calculate the production  rate of axions from the process
$\gamma\gamma\to     a$  in  a thermal   medium,    we approximate the
Bose-Einstein photon distribution by a Maxwell-Boltzmann one, i.e.\ we
use $e^{-\omega/T}$ instead  of  $1/(e^{\omega/T}-1)$ for   the photon
occupation number.  This    approximation is justified since   we  are
interested only in  axion masses and thus  axion energies of order the
temperature or larger. The production rate of axions of energy $E$ per
unit volume and unit energy interval is then found to be
\begin{equation}
\frac{d N_a}{dE}\ =\ \frac{g_{a\gamma\gamma}^2m^4}{128\,\pi^3}\,
p\,e^{-E/T}\,,
\end{equation}
where again $p=\sqrt{E^2-m^2}$ is the axion momentum. Integrating this
expression over   a standard solar model  we  find the  axion  flux at
Earth. It is approximately represented by
\begin{equation}
\Phi_a\ =\ 
1.68\times10^{9}~{\rm cm^{-2}~s^{-1}~keV^{-1}}
\, \bigg(\frac{g_{a\gamma\gamma}}{10^{-10}\ {\rm GeV}^{-1}}\bigg)^2\,
m^4\,p\,\left(\frac{10}{0.2+E^2}+1+0.0006\,E^3\right)\,e^{-E},
\end{equation}
where again  $m$, $E$  and $p$ are  to be  taken in keV.   For $1~{\rm
  keV}<E<16~{\rm keV}$ the quality of the approximation is better than
5\%. Both lower  and higher energies are irrelevant  for our purposes. 
In  Fig.~\ref{f3} we  display numerical  estimates of  the flux  of KK
axions  at Earth  as a  function of  their energy  for  three selected
values of  axion mass: $m= 5$,  10 and 15~keV. A  direct comparison of
Fig.~\ref{f2} with  Fig.~\ref{f3} reveals that  the photon coalescence
process becomes more important than  the Primakoff one for the heavier
KK-axion modes.

\subsection{Axion limit from solar energy loss}

As a next step we consider the energy loss of the Sun as a function of
$g_{a\gamma\gamma}$. To this end  we  first calculate the solar  axion
luminosity as a function of the KK axion mass
\begin{equation}
L_a(m)\ =\ 4\pi d_\odot^2 \int_m^\infty dE\,E\,\Phi_a(E)
\end{equation}
for the two  processes. Then we need   to sum over  all  KK modes with
their  different masses.   Instead,  we integrate  over the density of
modes which is $R^\delta$,  where  $R$ is the compactification  radius
and $\delta$ the number  of  compactified dimensions.  Therefore,  the
axion luminosity is
\begin{equation}
L_a\ =\ \frac{2\pi^{\delta/2}}{\Gamma(\delta/2)}\,R^\delta
\int_0^\infty dm\,m^{\delta-1} L_a(m)\, ,
\end{equation}
where the first factor is the surface of the $\delta$ dimensional unit
sphere, i.e.\  2 for $\delta=1$,  $2\pi$ for $\delta=2$ and $4\pi$ for
$\delta=3$.  Numerically, we write the result in the form
\begin{equation}\label{eq:solaraxflux}
L_a\ =\ A\,L_\odot\,
\bigg(\frac{g_{a\gamma\gamma}}{10^{-10}\ {\rm GeV}^{-1}}\bigg)^2\,
\bigg(\frac{R}{{\rm keV}^{-1}}\bigg)^\delta
\end{equation}
where  $L_\odot$ is the  luminosity of the Sun and  the  values of the
coefficients $A$  for   the two  processes and   different  dimensions
$\delta$ are given in Table~\ref{tab:fluxcoefficients}.  It depends on
$\delta$ which of the processes is more important.

\begin{table}
\begin{center}
\caption{\label{tab:fluxcoefficients} Coefficients $A$ for 
Eq.~(\protect\ref{eq:solaraxflux}).}
\medskip
\begin{tabular}{llll}
\hline
\hline
&Primakoff&Coalescence&Sum\\
\hline
$\delta=1$&0.015&0.0033&0.018                 \\ 
$\delta=2$&0.12 &0.067 &0.19                 \\
$\delta=3$&0.99 &1.06  &2.1                  \\
\hline
\end{tabular}
\end{center}
\end{table}

Helioseismology implies that a novel energy-loss mechanism of the Sun
must not exceed something like $0.2\,L_\odot$ \cite{Schlattl}.
This limit translates into the constraint
\begin{equation}
\bigg(\frac{g_{a\gamma\gamma}}{10^{-10}\ {\rm GeV}^{-1}}\bigg)\,
\bigg(\frac{R}{{\rm keV}^{-1}}\bigg)^{\delta/2}
<\cases{3.3&for $\delta=1$,\cr
1.0&for $\delta=2$,\cr
0.31&for $\delta=3$.\cr}
\end{equation}
As an example we use the simplest setting of $\delta = n = 2$ large
extra dimensions, with $M_{\rm F}=100$~TeV and $R=10^3$~keV$^{-1}$,
leading to $g_{a\gamma\gamma}<10^{-13}~{\rm GeV}^{-1}$.  For $\delta =
n = 3$ large extra dimensions, with $M_{\rm F}=1$ TeV and
$R=10^2$~keV$^{-1}$, we get an even better limit of
$g_{a\gamma\gamma}<0.3\times10^{-13}~{\rm GeV}^{-1}$.  This is to be
compared with the solar PQ axion limit of
$g_{a\gamma\gamma}<10^{-9}~{\rm GeV}^{-1}$ \cite{Schlattl}. Of course,
the KK limits could have been estimated by simply scaling the standard
limit with the multiplicity of KK modes and observing that the maximum
allowed mass is a few keV before the solar flux gets suppressed by the
kinematic threshold.

\setcounter{equation}{0}
\section{Flux of decay photons}

\subsection{Numerical estimates}

The KK  axions emerging from the Sun  are neither  nonrelativistic nor
strongly  relativistic. The  average speed is   0.95 (in units  of the
speed of  light) for $m=1~{\rm keV}$,  0.79 for 3~keV, 0.66 for 5~keV,
0.57 for 7~keV, and 0.51 for  9~keV. Therefore, the decay photons will
have a considerable  angular spread relative to  the  direction of the
Sun. The event  rate  in a   detector  thus depends crucially  on  its
geometry. For our  simple estimate  we  will assume that the  detector
consists  of a  volume $V$, and   that any x-ray produced  within this
volume  will be detected  with  unit efficiency, independently of  its
direction.

\begin{figure}[p]
   \leavevmode
 \begin{center}
   \epsfxsize=16.0cm
    \epsffile[0 0 539 652]{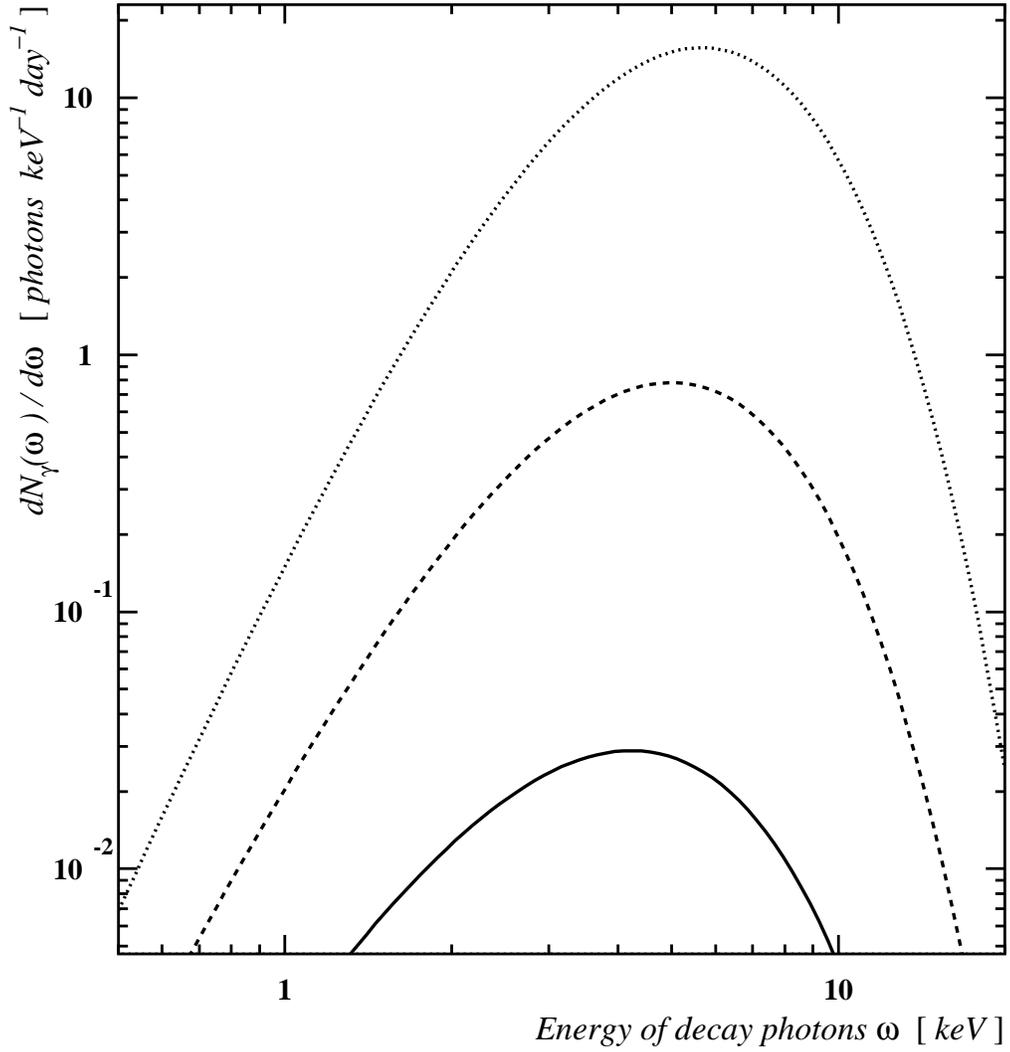}
 \end{center}
 \vspace{-0.8cm} 
\caption{Energy dependence of $d N_\gamma (\omega )/d \omega$, for
  $\delta = 1$ (solid line), $\delta = 2$ (dashed line) and $\delta =
  3$ (dotted line), with $g_{a\gamma\gamma} = 10^{-10}$ GeV$^{-1}$,
  $R=1$ keV$^{-1}$ and $V = 1$ m$^3$.}
\label{f1}
\end{figure}

In view of the solar energy-loss  limits derived above we further note
that even  keV-mass  axions are long-lived  relative  to the Sun-Earth
distance so  that    the axion flux    on its  way to     Earth is not
significantly diminished by radiative decays.  Therefore, at any given
time  the total number  of solar axions  of  mass $m$  per unit energy
interval in the detector is
\begin{equation}
\frac{d N_a}{dE}=\frac{V\Phi_a}{v}
\end{equation}
where  $v=p/E$ is the  axion  velocity. In  the  laboratory frame they
decay           with          a           rate                  $(m/E)
\Gamma_{a\to\gamma\gamma}=(g_{a\gamma\gamma}^2/64\pi)   m^4/E$,   each
decay  producing  2  photons   with  energies   which   are  uniformly
distributed in the   range  $(E-p)/2\leq  \omega\leq   (E+p)/2$.  This
implies that in order  to get a decay  photon  of energy $\omega$  the
parent axion must have $E\geq  \omega+m^2/4\omega$ and that the photon
energies  from a  given axion  decay  are  spread over an  interval of
length $p$.  Altogether, then, we find for the differential event rate
of decay photons from axions of mass $m$
\begin{equation}
\frac{d N_\gamma(m,\omega)}{d\omega}\ =\ 
\Gamma_{a\to\gamma\gamma}\, m\, V
\int_{\omega+m^2/4\omega}^\infty \!dE\, \frac{2\Phi_a}{p^2}\ .
\end{equation}
Finally, in order to obtain the total event rate due to all modes of
the tower of KK modes we proceed as before by integrating over the
density of modes so that
\begin{equation}
\frac{d N_\gamma(\omega)}{d\omega}\ =\ 
\frac{2\pi^{\delta/2}}{\Gamma(\delta/2)}\,R^\delta
V \int_0^\infty dm\,m^{\delta} 
\Gamma_{a\to\gamma\gamma}
\int_{\omega+m^2/4\omega}^\infty \!dE\, \frac{2\Phi_a}{p^2}\ .
\end{equation}
Numerically, we write this in the form
\begin{equation}\label{eq:rate}
\frac{d N_\gamma(\omega)}{d\omega}\ =\ 
A_\delta\,
\bigg(\frac{g_{a\gamma\gamma}}{10^{-10}\ {\rm GeV}^{-1}}\bigg)^4\,
\bigg(\frac{R}{{\rm keV}^{-1}}\bigg)^\delta
\bigg(\frac{V}{{\rm m}^{3}}\bigg)
f_\delta (\omega )
\end{equation}
where $A_\delta$ is a rate given in Table~\ref{tab:ratecoefficients}
and $f_\delta(\omega)$ is a spectrum with its integral normalized to
unity. These normalized functions are surprisingly well approximated
by the simple analytic form
\begin{equation}\label{eq:spectra}
f_\delta (\omega)\ =\ a\,\omega^b\,e^{-0.9\,\omega}
\end{equation}
where $a$ and $b$   are given in  Table~\ref{tab:ratecoefficients} for
each   $\delta$.   Of  course, $\omega$  is   understood   in keV.  In
Table~\ref{tab:ratecoefficients} we   also give  the  average   photon
energies. In  particular, as can    be seen from Fig.\  \ref{f1},  the
energy distributions  of the decay  photons are shifted towards to the
few keV energy range.

\begin{table}
\begin{center}
\caption{\label{tab:ratecoefficients} Coefficients for
the rate  
Eq.~(\protect\ref{eq:rate})
and for the spectra of 
Eq.~(\protect\ref{eq:spectra}).}
\medskip
\begin{tabular}{lllll}
\hline
\hline
&$A_\delta$ [day$^{-1}$]&$a$ [keV$^{-1}$]&$b$&$\langle\omega\rangle$ [keV]\\
\hline
$\delta=1$&0.16&0.0338&3.8&5.3         \\ 
$\delta=2$&4.7 &0.0107&4.5&6.1         \\
$\delta=3$&100.&0.0037&5.1&6.8         \\
\hline
\end{tabular}
\end{center}
\end{table}

\subsection{Experimental sensitivity}

On the  experimental side, we assume  a 1~m$^3$ cubic  detector of the
Micromegas type. This is a new kind of gas  detector which can be used
to  measure photon interactions with good  space and energy resolution
\cite{Micro}.  A  small detector of this kind,  with a surface  of $15
\times 15\ {\rm   cm^2}$, was used in   Saclay  (on the surface)   and
measured 1.2   neutral particles per second  in  a  1 keV  wide energy
interval centred at 1 keV. At these  energies, practically all photons
entering the chamber interact in the gas,  so we have a measurement of
the neutral  particle flux  through a   surface of  $15\times 15\ {\rm
  cm}^2$, which  is about $53\  {\rm neutral\ particles} /  {\rm m}^2/
{\rm sec}$ \cite{YG}.

In  the  search  for axion  decays  into   two  gammas, the background
originates  from two neutral particles interacting  in  the gas within
the resolving time of the chamber.  Therefore,  one can choose the gas
so  that the  mean   absorption length of   1  keV photons is  0.3  cm
\cite{PDG}.  As  a result, the  interaction points of the  two photons
from axion decay  will be very close  to each  other,  in a  cell with
volume $\Delta x  \Delta   y \Delta z   =  1\  {\rm cm}^3$.  In    the
Micromegas chamber $\Delta x$ and $\Delta y$ are measured directly and
$\Delta z$ is measured from the time interval between the two signals.
For $\Delta z = 1\ {\rm cm}$, the  time interval is $2 \times 10^{-7}$
sec.   Thus, in a small  cell of 1 cm$^3$  volume,  the rate of events
from  two uncorrelated  neutral  particles  is  $5.6 \times  10^{-12}$
events/sec.  As there  are $10^6$ cells when  going from 1 cm$^3$ cell
size to 1 m$^3$ size of the  detector, the background rate becomes 0.5
events per day.

At this point,  we should remark that  we have not used two additional
criteria to reduce the background:

\begin{itemize} 
  
\item[a)] Real photons in the   keV region entering the detector  from
  outside will  interact  very  close to  the  detector  walls. If one
  requires that the events occur in a fiducial volume at some distance
  (few cm) from the walls, only photons  generated inside the detector
  volume are important.

\item[b)] If axions are non-relativistic, the two photons will have
approximately equal energies.

\end{itemize}
Of   course, a  more  precise estimate   of  the background requires a
measurement  with a realistic   detector in the environment  where the
experiment is going to be performed.

Applying now Eq.~(\ref{eq:rate}) to the simplest  setting of $\delta =
n =   2$  large  extra  dimensions,   with $M_{\rm   F}=100$~TeV   and
$R=10^3$~keV$^{-1}$, we find the rate
\begin{equation}
  \label{Rexp1}
R_\gamma\ \approx\  0.05~{\rm events~day^{-1}~m^{-3}}~
\bigg(\frac{g_{a\gamma\gamma}}{10^{-12}\
  {\rm GeV}^{-1}}\bigg)^4\, .  
\end{equation}
Consequently, the suggested terrestrial  detector outlined  above will
be     sensitive     to    an  effective      $a\gamma\gamma$-coupling
$g_{a\gamma\gamma}    \stackrel{<}{{}_\sim}    2.\times      10^{-12}$
GeV$^{-1}$, corresponding to  a fundamental PQ mass $m_{\rm PQ}\approx
10^{-2}$  eV.   In particular,  for  $\delta =  n   =  3$  large extra
dimensions, with $M_{\rm  F}=1$ TeV and $R=10^2$~keV$^{-1}$, we obtain
an estimate for the rate
\begin{equation}
  \label{Rexp2}
R_\gamma\ \approx\  1.0~{\rm events~day^{-1}~m^{-3}}~
\bigg(\frac{g_{a\gamma\gamma}}{10^{-12}\ {\rm GeV}^{-1}}\bigg)^4\, .  
\end{equation}
{}From this last  result, one can  readily see that the axion detector
will  be maximally sensitive  to an effective $a\gamma\gamma$-coupling
$g_{a\gamma\gamma}    \stackrel{<}{{}_\sim}    6.\times      10^{-13}$
GeV$^{-1}$, corresponding to a fundamental  PQ mass $m_{\rm PQ}\approx
3.\times 10^{-3}$ eV.

Finally,    it would be interesting  to   know whether measurements of
$\gamma$-rays  coming from the  Sun could impose severe constraints on
the $2\gamma$-decay mode of axions and hence  on the parameters of the
higher-dimensional   axionic models  under  consideration  \cite{GR}.  
According to recent analyses  \cite{PF}, the solar x-ray luminosity in
the range of interest to us, i.e.\ above 0.4~keV, is
\begin{equation}
  \label{xray}
L_{\rm x-rays}\ \approx\ 10^9\ {\rm events}/{\rm cm}^2/{\rm sec}\
\approx\ 10^{17}\ {\rm events}/{\rm day}/{\rm m}^2\, .
\end{equation}
As the decay path  available for solar axions is  the distance to  the
Sun of  $1.5\times10^{11}~{\rm m}$, the   x-ray luminosity is  by many
orders of magnitude  larger than the one expected  from  the decays of
the KK axions.

\subsection{Laboratory limits on solar axions}

Lowest  level underground  experiments~\cite{AM} searching  for weakly
interacting  massive  particles  (WIMPs)  and  other  particles  offer
independent   limits  on   the   effective  axion-to-photon   coupling
$g_{a\gamma\gamma}$.   Specifically,   these  experiments  report  the
following lower limit on the integrated event rate in the energy range
below 10~keV:
\begin{equation}
  \label{Rexp}
R^{\rm exp}_\gamma\ \stackrel{<}{{}_\sim}\ 
20000~{\rm events~day^{-1}~m^{-3}}\,.    
\end{equation}
These highly  sensitive experiments  measure the deposited  energy but
they are unable to distinguish between 1-prong and 2-prong events.

Applying Eq.~(\ref{Rexp}) to  Eqs.~(\ref{Rexp1}) and (\ref{Rexp2}), we
are able  to derive  {\em for the  first time} experimental  limits on
$g_{a\gamma\gamma}$ in theories  with KK axions. In this  way, we find
the upper limits
\begin{equation}
  \label{ga1}
g_{a\gamma\gamma}\ \stackrel{<}{{}_\sim}\ 2.5\times 10^{-11}~{\rm GeV}^{-1},
\end{equation}
for $\delta = 2$ and $M_F = 100$~TeV, and 
\begin{equation}
  \label{ga2}
g_{a\gamma\gamma}\ \stackrel{<}{{}_\sim}\ 1.2\times 10^{-11}~{\rm GeV}^{-1},
\end{equation}
for  $\delta  =  3$  and  $M_F =  1$~TeV.   Evidently,  our  suggested
underground detector will  improve at least by one  order of magnitude
the present  experimental limits which we  derived in Eqs.~(\ref{ga1})
and  (\ref{ga2}). The latter  upper limits  should also  be contrasted
with the weaker  upper bound: $g_{a\gamma\gamma} \stackrel{<}{{}_\sim}
6.\times   10^{-10}$~GeV$^{-1}$,  which   is   obtained  from   recent
experimental searches  for conventional PQ axions coming  from the Sun
\cite{SM}.

\setcounter{equation}{0}
\section{Conclusions}

We  have examined the  potential of  an underground  detector shielded
from cosmic-ray  backgrounds for detecting  KK axions coming  from the
Sun.  The  solar KK axions may  be produced via  the Primakoff process
$\gamma + Ze \to Ze + a$ or via the photon coalescence process $\gamma
\gamma \to a$. In either case, we have calculated the expected flux of
the  KK axions,  as well  as  estimated possible  limits derived  from
helioseismology. We find that solar KK axions might lead to observable
signatures  in terrestrial experiments.   In fact,  the characteristic
2$\gamma$-decay mode of  the KK axions offers a  unique possibility to
drastically  reduce the cosmic  background by  coincidental triggering
both of the emitted photons.   Our elaborate estimates have shown that
a  terrestrial  detector  of  1  m$^3$  size may  be  sensitive  to  a
fundamental PQ-axion mass up to  $10^{-2}$ eV, which amounts to having
an  effective   axion-photon  coupling  $g_{a\gamma   \gamma}  \approx
2.\times  10^{-12}$  GeV$^{-1}$,  in   theories  with  2  large  extra
dimensions  and a  fundamental quantum-gravity  scale  $M_{\rm F}=100$
TeV.  In particular, in theories  with 3 large compact dimensions with
$M_{\rm  F}=1$  TeV, the  suggested  detector  is  capable of  probing
PQ-axion  masses up  to  $3.\times 10^{-3}$  eV,  corresponding to  an
effective axion-photon  coupling $g_{a\gamma \gamma}  \approx 6.\times
10^{-13}$  GeV$^{-1}$.  Most  importantly,  the experimental  detector
under discussion will  considerably improve, at least by  one order of
magnitude,  the  corresponding   experimental  limits  on  $g_{a\gamma
  \gamma}$  in   theories  with  KK   axions,  which  we   derived  in
Eqs.~(\ref{ga1}) and  (\ref{ga2}) based on present  data obtained from
underground experiments.

\end{document}